\documentclass[twocolumn,showpacs,aps,prl,superscriptaddress]{revtex4}
\usepackage{epsf,psfig}

\begin{document}

\title{

Scaling Properties of Hyperon Production in Au+Au Collisions at
$\sqrt{s_{NN}}$ = 200 GeV}
\date{\today}

\affiliation{}
\affiliation{Argonne National Laboratory, Argonne, Illinois 60439}
\affiliation{University of Birmingham, Birmingham, United Kingdom}
\affiliation{Brookhaven National Laboratory, Upton, New York 11973}
\affiliation{California Institute of Technology, Pasadena, California 91125}
\affiliation{University of California, Berkeley, California 94720}
\affiliation{University of California, Davis, California 95616}
\affiliation{University of California, Los Angeles, California 90095}
\affiliation{Carnegie Mellon University, Pittsburgh, Pennsylvania 15213}
\affiliation{Creighton University, Omaha, Nebraska 68178}
\affiliation{Nuclear Physics Institute AS CR, 250 68 \v{R}e\v{z}/Prague, Czech Republic}
\affiliation{Laboratory for High Energy (JINR), Dubna, Russia}
\affiliation{Particle Physics Laboratory (JINR), Dubna, Russia}
\affiliation{University of Frankfurt, Frankfurt, Germany}
\affiliation{Institute of Physics, Bhubaneswar 751005, India}
\affiliation{Indian Institute of Technology, Mumbai, India}
\affiliation{Indiana University, Bloomington, Indiana 47408}
\affiliation{Institut de Recherches Subatomiques, Strasbourg, France}
\affiliation{University of Jammu, Jammu 180001, India}
\affiliation{Kent State University, Kent, Ohio 44242}
\affiliation{Institute of Modern Physics, Lanzhou, China}
\affiliation{Lawrence Berkeley National Laboratory, Berkeley, California 94720}
\affiliation{Massachusetts Institute of Technology, Cambridge, MA 02139-4307}
\affiliation{Max-Planck-Institut f\"ur Physik, Munich, Germany}
\affiliation{Michigan State University, East Lansing, Michigan 48824}
\affiliation{Moscow Engineering Physics Institute, Moscow Russia}
\affiliation{City College of New York, New York City, New York 10031}
\affiliation{NIKHEF and Utrecht University, Amsterdam, The Netherlands}
\affiliation{Ohio State University, Columbus, Ohio 43210}
\affiliation{Panjab University, Chandigarh 160014, India}
\affiliation{Pennsylvania State University, University Park, Pennsylvania 16802}
\affiliation{Institute of High Energy Physics, Protvino, Russia}
\affiliation{Purdue University, West Lafayette, Indiana 47907}
\affiliation{Pusan National University, Pusan, Republic of Korea}
\affiliation{University of Rajasthan, Jaipur 302004, India}
\affiliation{Rice University, Houston, Texas 77251}
\affiliation{Universidade de Sao Paulo, Sao Paulo, Brazil}
\affiliation{University of Science \& Technology of China, Hefei 230026, China}
\affiliation{Shanghai Institute of Applied Physics, Shanghai 201800, China}
\affiliation{SUBATECH, Nantes, France}
\affiliation{Texas A\&M University, College Station, Texas 77843}
\affiliation{University of Texas, Austin, Texas 78712}
\affiliation{Tsinghua University, Beijing 100084, China}
\affiliation{Valparaiso University, Valparaiso, Indiana 46383}
\affiliation{Variable Energy Cyclotron Centre, Kolkata 700064, India}
\affiliation{Warsaw University of Technology, Warsaw, Poland}
\affiliation{University of Washington, Seattle, Washington 98195}
\affiliation{Wayne State University, Detroit, Michigan 48201}
\affiliation{Institute of Particle Physics, CCNU (HZNU), Wuhan 430079, China}
\affiliation{Yale University, New Haven, Connecticut 06520}
\affiliation{University of Zagreb, Zagreb, HR-10002, Croatia}

\author{J.~Adams}\affiliation{University of Birmingham, Birmingham, United Kingdom}
\author{M.M.~Aggarwal}\affiliation{Panjab University, Chandigarh 160014, India}
\author{Z.~Ahammed}\affiliation{Variable Energy Cyclotron Centre, Kolkata 700064, India}
\author{J.~Amonett}\affiliation{Kent State University, Kent, Ohio 44242}
\author{B.D.~Anderson}\affiliation{Kent State University, Kent, Ohio 44242}
\author{M.~Anderson}\affiliation{University of California, Davis, California 95616}
\author{D.~Arkhipkin}\affiliation{Particle Physics Laboratory (JINR), Dubna, Russia}
\author{G.S.~Averichev}\affiliation{Laboratory for High Energy (JINR), Dubna, Russia}
\author{Y.~Bai}\affiliation{NIKHEF and Utrecht University, Amsterdam, The Netherlands}
\author{J.~Balewski}\affiliation{Indiana University, Bloomington, Indiana 47408}
% ...

\collaboration{STAR Collaboration}\noaffiliation

\begin{abstract}
We present the scaling properties of $\Lambda$, $\Xi$, $\Omega$ and
their anti-particles produced at mid-rapidity in $Au+Au$ collisions at
RHIC at $\sqrt{s_{NN}}$ = 200 GeV.  The yield of multi-strange baryons
per participant nucleon increases from peripheral to central
collisions more rapidly than the $\Lambda$ yield, which appears to
correspond to an increasing strange quark density of matter produced.
The value of the strange phase space occupancy factor $\gamma_{s}$, obtained
from a thermal model fit to the data, approaches unity for the most central
collisions. We also show that the nuclear modification factors, $R_{CP}$,
of $\Lambda$ and $\Xi$ are consistent with each other and with that of
protons in the transverse momentum range 2.0 $< p_T <$ 5.0 GeV/c.
This scaling behaviour is consistent with a scenario of hadron formation 
from constituent quark degrees of freedom through quark recombination or
coalescence.
\end{abstract}


\maketitle

Lattice Quantum ChromoDynamics calculations predict that a new state
of matter, the Quark Gluon Plasma (QGP), can be formed at zero baryon
density in nuclear collisions when the temperature exceeds $160-170$
MeV~\cite{LQCD}.  Strange quarks, whose mass is comparable to the
critical temperature, are expected to be abundantly produced by
thermal parton interactions in the high temperature QGP phase.  Due to
the corresponding increase in the strange quark density, hyperon
production is expected to be enhanced in high energy nuclear
collisions, the enhancement increasing with the number of strange
valence quarks in the hyperon~\cite{RafMuller}.  Such an effect has
already been observed in various fixed-target experiments at lower
energy by comparing the number of hyperons produced per participating
nucleon in nucleus-nucleus and proton-nucleus
collisions~\cite{WA97,NA49,NA57}.  In this letter, we study the
centrality dependence of hyperon production in $Au+Au$ collisions at a
collision energy of $\sqrt{s_{NN}}$ = 200 GeV, which is an order of
magnitude higher than that previously achieved. We also study the
transverse momentum dependence of hyperon production in central and
peripheral collisions in an attempt to shed light upon the possible
production mechanisms.

Previous studies have shown that ratios of hadron yields in high
energy nucleus-nucleus collisions are generally well described by
statistical models in the grand canonical limit, where baryo-chemical
potential and temperature are parameters~\cite{BMS, RL, BecaFerr}.  A
strangeness phase-space occupancy factor, $\gamma_s$, is sometimes
introduced to describe the extent to which strangeness reaches its
equilibrium abundance. In this framework, the amount of strangeness
produced per participating nucleon ($N_{part}$) is directly related to
the value of $\gamma_s$.  The centrality dependence of $\gamma_s$
therefore provides a quantitative measure of strangeness equilibration
as a function of system size in nucleus-nucleus
collisions~\cite{CleyQM}, under the assumption that the grand
canonical approximation remains valid in non-central collisions.

By contrast, at high transverse momentum, hadrons are thought to be
produced via incoherent hard scatterings, which, in the absence of any
nuclear medium effects, should scale with the number of binary
nucleon-nucleon collisions
($N_{binary}$)~\cite{Star7,Glaub}. Measurements of hadron production
in $Au+Au$ collisions at RHIC have shown that not only is there a
deviation from binary scaling in central collisions~\cite{Phe1,Star7},
but also a distinct difference in the scaling behaviour of baryons and
mesons in the intermediate transverse momentum range, 2 $< p_{T} <$ 5
GeV/c~\cite{Phe2,Star2}. A strong particle-type dependence is not
predicted by conventional Monte Carlo (MC) event simulators such as
HIJING, where hadron formation in this region is dominated by
independent parton fragmentation~\cite{LUND}.  On the other hand,
quark recombination (coalescence) models have been successful in
explaining the observed deviation from binary scaling for baryons and
mesons in central collisions~\cite{MV,LK2,FMNB,GKL}, as well as
providing an explanation for the particle-type dependence of measured
azimuthal anisotropies at intermediate $p_{T}$ in non-central
collisions~\cite{Star2}.  By extending these previous studies to
include multi-strange baryons we provide a more stringent test of
recombination models. Furthermore, it may allow us to probe the
differences between strange and light (up and down) quark
distributions produced in nucleus-nucleus collisions.

The STAR Time Projection Chamber (TPC) measures the trajectories and
momenta of charged particles produced in each collision in the
pseudo-rapidity range $\left| \eta\right| < 1.8$~\cite{Star3}. The
detector operates within a solenoidal magnetic field of $0.5$ Tesla
whose axis is aligned with the beam. A central trigger barrel,
covering the pseudo-rapidity region $\left| \eta\right| < 1$, and two
zero-degree calorimeters are used as trigger detectors. A total of
$1.6\times10^6$ minimum-bias trigger collisions and $1.5\times 10^6$
central trigger collisions were used for this analysis.  A detailed
description of the analysis including particle reconstruction, track
quality, decay vertex topology cuts and calculation of the detection
efficiency can be found elsewhere~\cite{Star1,Star4,Star8}.
In this study $\Lambda(\overline{\Lambda})$,
$\Xi^{-}(\overline{\Xi}^{+})$ and $\Omega^{-}(\overline{\Omega}^{+})$
have been measured in rapidity intervals of $|y|<1$, $0.75$ and
$0.75$, respectively. In order to increase statistics, the results for
$\Omega^{-}$ and $\overline{\Omega}^{+}$ have been combined.  Within
the chosen rapidity intervals the particle reconstruction efficiency
is a function of transverse momentum and lifetime.  The efficiency
calculations were based on the probability of finding Monte Carlo
generated particles after processing them through a TPC detector
response simulation, embedding them into real events and then
reconstructing them as real data. The collision centrality was defined
by the charged particle multiplicity measured in the TPC in the
pseudo-rapidity range $\left| \eta\right| < 0.5$. Five centrality bins
were selected corresponding to the following ranges in the total
hadronic cross section ($0-5\%$, $10-20\%$, $20-40\%$, $40-60\%$,
$60-80\%$). The $0-5\%$ bin represents the most central collisions and
was obtained from the central trigger sample. The remaining bins were
obtained from the minimum-bias sample.  Due to relatively poor
statistics, the 5-10\% bin and the $\Omega$ 10-20\% and 60-80\% bins
were omitted from this analysis.

\begin{figure}
\hspace*{-1cm}
\centering\mbox{
\psfig{figure=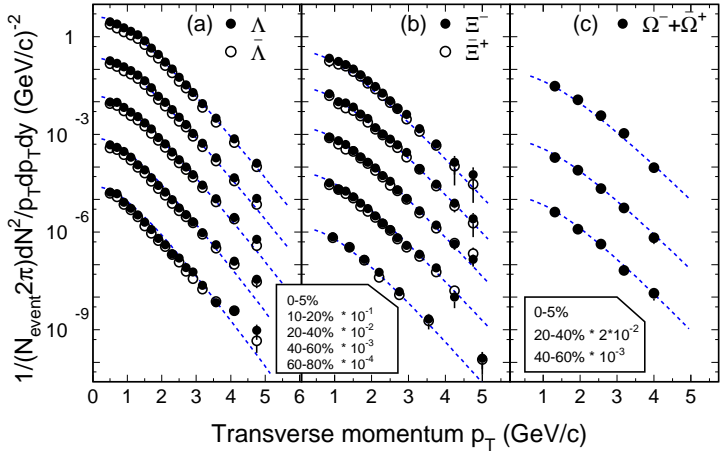,height=6.7cm}}
\vspace{-0.5cm}
\vspace*{-0.5cm}
\caption{Transverse momentum distributions of (a)
$\Lambda$($\overline{\Lambda}$) for $|y|<1.0$, (b) $\Xi^{-}$
($\overline{\Xi}^{+}$) for $|y|<0.75$ and (c)
$\Omega^{-}+\overline{\Omega}^{+}$ for $|y|<0.75$ in $Au+Au$ collisions
at $\sqrt{s_{NN}} = 200$ GeV as a function of centrality. The
$\Lambda$ spectra were corrected for weak decay of $\Xi$, $\Xi^{0}$
and $\Omega$. Scale factors were applied to the spectra for
clarity. Only statistical errors are shown. The dashed curves show a
Boltzmann fit to the $\Lambda$, $\Xi^{-}$ and
$\Omega^{-}+\overline{\Omega}^{+}$ data, the fits to the
$\overline{\Lambda}$ and $\overline{\Xi}^{+}$ are omitted for
clarity. }
\vspace{-0.5cm}
\label{fig:spectra}
\end{figure}

Figure~\ref{fig:spectra} shows the transverse momentum distributions
of $\Lambda$($\overline{\Lambda}$), $\Xi^{-}$($\overline{\Xi}^{+}$) and
$\Omega^{-}+\overline{\Omega}^{+}$ measured at mid-rapidity and as
function of centrality. The errors shown on the data points are statistical
only. The $\Lambda$ spectra were corrected for feed-down from
multi-strange baryon weak decays, based upon the measured $\Xi$
 and $\Omega$ spectra. The feed-down correction depends sensitively on both
experimental acceptance and the cuts used in the analysis. The
contribution to the $\Lambda$ spectrum from $\Xi$ and $\Omega$ decays
is at the 15\% level. The feed-down contribution to the $\Xi$ spectrum
from $\Omega$ decays is negligible. The measured $p_{T}$ coverage is
about 70\% for $\Lambda$ and 60\% for $\Xi$ and $\Omega$. The
total integrated yields ($dN/dy$) were extracted from Boltzmann fits
to the spectra and are presented in Table ~\ref{tab:dNdy}.

\begin{table*}
\begin{ruledtabular}
\begin{tabular}{lccccc}
Centrality  & 0--5$\%$ & 10--20$\%$ & 20--40$\%$ & 40--60$\%$ & 60--80$\%$ \\ 
\hline
$<N_{part}>$ & 352$\pm$3 & 235$\pm$9 & 141$\pm$8 & 62$\pm$9 & 21$\pm$6\\
\hline
$\Lambda$ & 16.7$\pm$0.2$\pm$1.1 & 10.0$\pm$0.1$\pm$0.7 & 5.53$\pm$0.05$\pm$0.39 & 2.07$\pm$0.03$\pm$0.14 & 0.58$\pm$0.01$\pm$0.04 \\ 
          & 309$\pm$1$\pm$8 & 308$\pm$1$\pm$8 & 303$\pm$1$\pm$8 & 297$\pm$2$\pm$10 & 287$\pm$3$\pm$10 \\ 
\hline
$\overline{\Lambda}$ &  12.7$\pm$0.2$\pm$0.9 & 7.7$\pm$0.1$\pm$0.5 & 4.30$\pm$0.04$\pm$0.30 & 1.64$\pm$0.03$\pm$0.11 & 0.48$\pm$0.01$\pm$0.03\\  
          & 310$\pm$1$\pm$7 & 309$\pm$1$\pm$8 & 306$\pm$1$\pm$9 & 298$\pm$2$\pm$10 & 282$\pm$3$\pm$10 \\ 
\hline
$\Xi^{-}$ & 2.17$\pm$0.06$\pm$0.19 & 1.41$\pm$0.04$\pm$0.08 & 0.72$\pm$0.02$\pm$0.02 & 0.26$\pm$0.01$\pm$0.02 & 0.063$\pm$0.004$\pm$0.003 \\ 
          & 335$\pm$4$\pm$7 & 331$\pm$4$\pm$8 & 326$\pm$3$\pm$6 & 325$\pm$4$\pm$7 & 320$\pm$8$\pm$13 \\ 
\hline
$\overline{\Xi}^{+}$ & 1.83$\pm$0.05$\pm$0.20 & 1.14$\pm$0.04$\pm$0.08 & 0.62$\pm$0.02$\pm$0.03 & 0.23$\pm$0.01$\pm$0.02 & 0.061$\pm$0.004$\pm$0.002 \\ 
                     & 335$\pm$4$\pm$9 & 334$\pm$4$\pm$9 & 327$\pm$3$\pm$6 & 327$\pm$5$\pm$7 & 302$\pm$8$\pm$16 \\ 
\hline
$\Omega+\overline{\Omega}^{+}$ & 0.53$\pm$0.04$\pm$0.04 & - & 0.17$\pm$0.02$\pm$0.01 & 0.063$\pm$0.008$\pm$0.004 & - \\
                               & 353$\pm$9$\pm$10 & - & 348$\pm$15$\pm$12 & 336$\pm$17$\pm$13 & - \\
\end{tabular}
\caption{Integrated yields $dN/dy$ and inverse slope parameters $T$ (MeV) extracted from a Boltzmann fit to the $p_{T}$ spectra of $\Lambda$$(\overline{\Lambda})$, $\Xi^{-}$$(\overline{\Xi})^{+}$ and
$\Omega^{-}+\overline{\Omega}^{+}$ at mid-rapidity. Statistical and
systematic errors are presented.  Also shown for each centrality is
$<N_{part}>$, the number of participants, extracted from a Monte Carlo
Glauber model calculation~\cite{Star7,Glaub}.}
\label{tab:dNdy}
\end{ruledtabular}
\end{table*}

The systematic error on the reconstructed yields was studied as a
function of $p_T$. Three main factors contribute to the systematic
error: (i) subtle differences between the Monte Carlo simulation and
real data, which make the reconstructed yields sensitive to the choice
of geometric cuts used to improve the signal to background ratio, (ii)
sensitivity to the method used to subtract the remaining background
after geometric cuts have been applied, and (iii) measured differences
in the yield dependent on the direction of the applied magnetic field.
At low $p_T$, the dominant contribution to the systematic error is due
to the choice of cuts. Here, the systematic error was estimated by
varying the cuts about their optimal values and observing the change
in the reconstructed yield. At high $p_T$ the systematic error is
dominated by the differences observed in the reconstructed yield for
the two magnetic field settings.  In order to determine the systematic
uncertainty on the total yield and the inverse slope parameter for
each particle and centrality class, $p_T$ dependent systematic errors
were added to the data points shown in figure~\ref{fig:spectra} and
included in a second fit.  The systematic errors shown in
Table~\ref{tab:dNdy} reflect the difference between the two fits.  We
also investigated the choice of function used to fit the data.
Although the Boltzmann function gave a better fit, an exponential
function could not be excluded.  Exponential fits to the data gave a
5-6\% higher yield on average and a larger inverse slope parameter by
40-50 MeV. These differences are not included in the errors shown in
Table~\ref{tab:dNdy}.
\begin{figure}[t]
\centering\mbox{
\psfig{figure=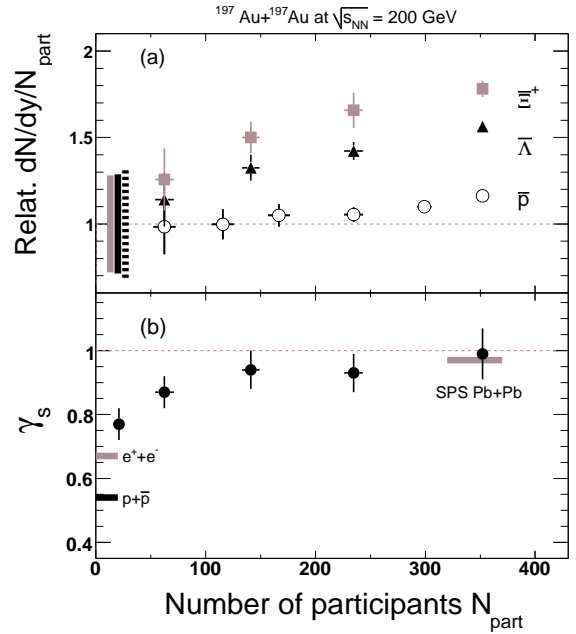,width=8.2cm}} 
\vspace{-0.2cm}
\caption{(a) The corrected integrated yield $dN/dy$ at mid-rapidity for
$\overline{\Xi}^{+}$, $\overline{\Lambda}$ and $\bar{p}$ divided by
$N_{part}$, normalized to the most peripheral centrality interval
($60-80\%$), plotted as a function of $N_{part}$. The gray, black and
dashed bands represent the errors on the normalization to the most
peripheral bin for the $\bar{\Xi}^+$, $\bar{\Lambda}$ and $\bar{p}$.
Other errors shown are statistical only.  (b) $\gamma_{s}$ as a
function of $N_{part}$ calculated from thermal model fits to the
measured particle yields ($\pi$,K,p~\cite{kpp}, $\Lambda,\Xi,\Omega$
and their anti-particles) at 200 GeV. Values for $e^++e^-$ and p+\=p
collisions at $\sqrt{s_{NN}}$ = 91 and 200 GeV respectively and for
Pb+Pb SPS collisions at $\sqrt{s_{NN}}$ = 17.2 GeV are shown for
comparison~\cite{FBec1,FBecpp,FBec}.}
\vspace{-0.5cm}
\label{fig:npart}
\end{figure}

Figure~\ref{fig:npart}(a) presents the strange anti-particle yields,
$dN/dy$, divided by $N_{part}$. For clarity, only statistical errors
are shown. All data points are normalized to the values obtained in
the most peripheral collisions (centrality bin $60-80\%$). The
centrality dependence of the anti-proton yield is also shown for
comparison~\cite{kpp}.  Strange anti-particles are chosen because all
valence quarks must have been created in the collision, although
similar results are also obtained for strange particles.  In a
geometrical description of nuclear collisions the number of
participant nucleons is proportional to the initial overlapping volume
of the colliding nuclei. The integrated yield is dominated by the low
$p_{T}$ region where particle production originates from mainly soft
(non-perturbative) processes. The integrated yield per participating
nucleon may be a measure of the formation probability of a hadron from
the bulk. As such, we would expect it to be sensitive to the density
of the hadron's constituent quarks in the system.  We note that there
appears to be a hierachy of particle production dependent upon
strange-quark content, which has also been observed at lower
energies~\cite{WA97,NA57}.  This may reflect an increase in the
strange quark density in more central collisions.

Thermal-statistical models have been very successful in describing
particle yields in various systems at different
energies~\cite{BMS,RL}.  Within such models, the densities of strange
particles, including strange resonances, are governed by statistical
laws.  The possible non-equilibrium of strange quarks is taken into
account by introducing a phase space occupancy factor, $\gamma_s$.
With the measured yields of strange baryons and other hadrons, such as
pions, kaons, protons and their anti-particles~\cite{kpp}, we have
performed a fit using the statistical model described in~\cite{Cley}
to determine $\gamma_{s}$ as a function of the number of participants,
as shown in figure~\ref{fig:npart}(b). We find that the value of
$\gamma_{s}$ increases from about $0.8$ in peripheral collisions to
about $1.0$ in central collisions. In each case we obtained a
freeze-out temperature around 165~MeV.  According to the model, the
$\Lambda$ yield depends linearly on $\gamma_{s}$ while the yield of
$\Xi$ depends on $\gamma_{s}^2$.  This is consistent with behavior
observed in figure~\ref{fig:npart}(a).  The fact that $\gamma_{s}$
approaches unity when $N_{part} >$ 150 suggests that the strange quark
abundance tends to equilibrate as the system-size increases.  A recent
analysis of hadron yields in nucleus-nucleus collisions at
$\sqrt{s_{NN}}$~=~17.2~GeV, using a different thermal model, also
found that $\gamma_{s}$ approaches unity at mid-rapidity in central
collisions~\cite{FBec}, whereas statistical analyses of elementary
e$^{+}$+e$^{-}$ and p+\=p collisions at various energies yield a value
of $\gamma_{s}$ significantly less than unity~\cite{FBec1,FBecpp}.

We studied the effect of including different combinations of particles
in the fit and found that particle ratios involving protons and
$\Lambda$ are important in constraining the freeze-out temperature and
$\gamma_{s}$, respectively. The value and centrality dependence of
$\gamma_{s}$ is relatively insensitive to the inclusion of other
particle ratios in the fit. The errors shown in
figure~\ref{fig:npart}(b) reflect the variation of $\gamma_s$ found in
this study.

\begin{figure}[b]
\centering\mbox{
\psfig{figure=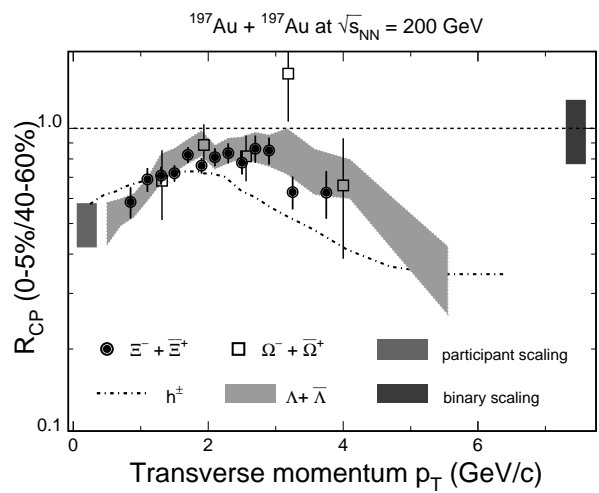,height=6.8cm}} 
\vspace{-0.5cm}
\caption{$R_{CP}$ for $\Xi^{-}+\overline{\Xi}^{+}$ and
 $\Omega^{-}+\overline{\Omega}^{+}$ at mid-rapidity (centrality
 interval : $0-5\%$ vs. $40-60\%$). A dashed line for charged hadrons
 and gray band for $\Lambda+\overline{\Lambda}$ are shown as
 comparison.  The gray rectangles represent participant and binary
 scalings.}
\vspace{-0.2cm}
 \label{fig:RCP} \end{figure}

In order to investigate the scaling behaviour of hyperon production in
the intermediate transverse momentum region, figure~\ref{fig:RCP}
shows the nuclear modification factor ($R_{CP}$)~\cite{Star2} for
$\Xi^{-}+\overline{\Xi}^{+}$ and $\Omega+\overline{\Omega}$.  The
nuclear modification factor was found by forming the ratio of the
$p_T$ spectra of the $0-5\%$ and $40-60\%$ centrality bins, after
normalising each spectrum to the average number of binary collisions,
appropriate for each centrality range, obtained from a Monte Carlo
Glauber calculation~\cite{Star7,Glaub}.  The $40-60\%$ centrality bin
was chosen as the reference because of the limited statistics of
$\Omega+\overline{\Omega}$ in the $60-80\%$ bin.  Also shown in
figure~\ref{fig:RCP} are the previously published results for charged
hadrons and $\Lambda+\overline{\Lambda}$ for the same centrality
bins~\cite{Star2}. The dark gray rectangular boxes on the plot
represent the expected $R_{CP}$ range for $N_{part}$ and $N_{bin}$
scalings, indicating the range of uncertainty in calculating the
number of participants and of binary collisions for each centrality.
Although the $p_T$ integrated yield per participating nucleon of $\Xi$
increases faster with $N_{part}$ than for $\Lambda$ hyperons, in the
interval $1.8 < p_{T} < 3.5$ GeV/$c$, the $p_{T}$ dependence of
$R_{CP}$ for $\Xi^{-}+\overline{\Xi}^{+}$ and
$\Omega^{-}+\overline{\Omega}^{+}$ are similar and coincide with the
trend previously shown for $\Lambda+\overline{\Lambda}$. The $R_{CP}$
of hyperons exhibits little suppression while mesons (approximated by
the dashed line) have a distinctly different trend.  The difference in
$R_{CP}$ for baryons and mesons in the intermediate $p_{T}$ region has
previously been discussed in the framework of recombination (or
coalescence) models~\cite{Star2,Phe1,FMNB,HwaYang}.  The results
presented here appear to confirm that the difference is dependent upon
the number of constituent quarks and is not a mass effect. Further
weight is given to this argument by a recent measurement of the
nuclear modification factor of protons~\cite{Phe2},
$K(892)^{*}$~\cite{Kstar} and $\phi$ mesons~\cite{jingguo}.  The
similarity between $\Lambda$ and $\Xi$ $R_{CP}$ at intermediate $p_T$
reinforces the notion of a baryon-meson difference. Futhermore, it
suggests that the strange quark distribution scales with centrality in
a similar way to up and down quarks, since baryons with different
strangeness content seem to follow the same pattern. This observation
is consistent with recent elliptic flow measurements of $\Lambda$,
$\Xi$ and $\Omega$ at intermediate $p_T$~\cite{200v2}.

In this letter, we have presented the scaling properties of strange
baryon production in $Au+Au$ collisions at $\sqrt{s_{NN}}$ = 200 GeV.
By studying the hyperon yields scaled by $N_{part}$ and the centrality
dependence of $\gamma_s$ within the framework of a thermal model, we
have found that strangeness equilibrium appears to have been achieved
in central collisions at RHIC.  We have also investigated the
centrality dependence of the transverse momentum distributions of
hyperons.  We find that hyperon yields in central collisions fall
below the expectation for binary scaling for $p_T > 3$ GeV/c and that
the nuclear modification factor $R_{CP}$ is similar for all hyperons
independent of their mass or strangeness content. In addition we note
that the $R_{CP}$ of hyperons is similar to that of protons, a feature
that is consistent with models of hadron formation based upon quark
recombination.

\vspace{0.2cm}
{\bf Acknowledgments:} We thank the RHIC Operations Group and RCF at
BNL, and the NERSC Center at LBNL for their support. This work was
supported in part by the HENP Divisions of the Office of Science of
the U.S.  DOE; the U.S. NSF; the BMBF of Germany; IN2P3, RA, RPL, and
EMN of France; EPSRC of the United Kingdom; FAPESP of Brazil; the
Russian Ministry of Science and Technology; the Ministry of Education
and the NNSFC of China; SFOM of the Czech Republic, FOM and UU of the
Netherlands, DAE, DST, and CSIR of the Government of India; the Swiss
NSF.
                                     
\end{document}